\documentclass[12pt,letterpaper]{article}
\pdfoutput=1

\usepackage{amsmath,amssymb,calc, latexsym, bbm, array, amsthm}
\usepackage{graphicx}
\usepackage{subfig}
\usepackage[nosort]{cite}
\usepackage{epsfig,psfrag}
\usepackage{slashed}

\usepackage{xcolor,colortbl}
\usepackage[colorlinks=true,linkcolor=black,citecolor=black,filecolor=black,urlcolor=black]{hyperref}

\definecolor{detail}{RGB}{110,110,110}

\newif\ifcomments
\commentstrue

\usepackage{tikz,amsmath}
\usetikzlibrary{shapes.geometric}
\def\weight #1 at (#2,#3){\node[fill=black, inner sep=1pt, label=40:{\small #1},circle] at (#2,#3) {}; }

\usepackage{ulem}
\usepackage{color}
\definecolor{burgundy}{rgb}{0.8,.2,.2}

\makeatletter
\renewcommand\section{\@startsection {section}{1}{\z@}%
                                   {-3.5ex \@plus -1ex \@minus -.2ex}
                                   {2.3ex \@plus.2ex}%
                                   {\normalfont\large\bfseries}}
\renewcommand\subsection{\@startsection{subsection}{2}{\z@}%
                                     {-3.25ex\@plus -1ex \@minus -.2ex}%
                                     {1.5ex \@plus .2ex}%
                                     {\normalfont\bfseries}}
\makeatother

\theoremstyle{plain}

\theoremstyle{definition}

\let\non\nonumber

\let\a=\alpha

\let\l=\lambda

\let\x=\xi
\let\y=\psi

\newcommand{\del}{\partial}

\def\one{^{(1)}}

\newcommand{\bea}{\begin{eqnarray}}
\newcommand{\eea}{\end{eqnarray}}
\newcommand{\be}{\begin{equation}}
\newcommand{\ee}{\end{equation}}
\def\ba#1\ea{\begin{align}#1\end{align}}
\def\ban#1\ean{\begin{align*}#1\end{align*}}
\newcommand{\bma}{\begin{pmatrix}}
\newcommand{\ema}{\end{pmatrix}}

\newcommand{\F}{{\mathbb F}}

\renewcommand{\O}{\operatorname{O}}

\newcommand{\La}{\Lambda}

\newcommand{\G}{\Gamma}

\newcommand{\dd}{\delta}

\newcommand{\f}{\psi}

\newcommand{\C}[1]{$(\ref{#1})$}

\def\IZ{\relax\ifmmode\mathchoice
{\hbox{\cmss Z\kern-.4em Z}}{\hbox{\cmss Z\kern-.4em Z}}
{\lower.9pt\hbox{\cmsss Z\kern-.4em Z}} {\lower1.2pt\hbox{\cmsss
Z\kern-.4em Z}}\else{\cmss Z\kern-.4em Z}\fi}
\def\IR{\relax{\rm I\kern-.18em R}}

\def\one{{\hbox{ 1\kern-.8mm l}}}

\def\tr{{\rm tr\,}}

\newlength{\bredde}
\def\slash#1{\settowidth{\bredde}{$#1$}\ifmmode\,\raisebox{.15ex}{/}
\hspace*{-\bredde} #1\else$\,\raisebox{.15ex}{/}\hspace*{-\bredde}
#1$\fi}

\newsavebox{\zzzbar}
\sbox{\zzzbar}
  {\setlength{\unitlength}{0.9em}
  \begin{picture}(0.6,0.7)
  \thinlines
  \put(0,0){\line(1,0){0.6}}
  \put(0,0.75){\line(1,0){0.575}}
  \multiput(0,0)(0.0125,0.025){30}{\rule{0.3pt}{0.3pt}}
  \multiput(0.2,0)(0.0125,0.025){30}{\rule{0.3pt}{0.3pt}}
  \put(0,0.75){\line(0,-1){0.15}}
  \put(0.015,0.75){\line(0,-1){0.1}}
  \put(0.03,0.75){\line(0,-1){0.075}}
  \put(0.045,0.75){\line(0,-1){0.05}}
  \put(0.05,0.75){\line(0,-1){0.025}}
  \put(0.6,0){\line(0,1){0.15}}
  \put(0.585,0){\line(0,1){0.1}}
  \put(0.57,0){\line(0,1){0.075}}
  \put(0.555,0){\line(0,1){0.05}}
  \put(0.55,0){\line(0,1){0.025}}
  \end{picture}}

\newcommand{\ket}[1]{|{#1}\rangle}

\newcommand{\ena}{\end{eqnarray}}
\newcommand{\beqa}{\begin{eqnarray}}
\newcommand{\eeqa}{\end{eqnarray}}

\def\G{\Gamma}

\def\d{{\rm d}}

\newfont{\goth}{ygoth.tfm scaled 1200}                   

\def\a{\alpha}

\def\th{\theta}
\def\f{\phi}

\def\j{\psi}

\def\l{\lambda}

\def\x{\xi}

\def\F{\Phi}
\def\G{\Gamma}
\def\J{\Psi}

\def\O{\Omega}
\renewcommand{\O}{{\mathcal{O}}}

 \numberwithin{equation}{section}

\def\1{{(1)}}
\def\2{{(2)}}
\def\3{{(3)}}

\def\1{{\bf 1}}
\def\a{{\alpha}}

\begin{document}
\begin{titlepage}

\begin{center}

{\Large \bf Marginal deformations of heterotic $G_2$ sigma models}

\vskip 1.25 cm {\bf Marc-Antoine Fiset$^{a\,}$\footnote{marc-antoine.fiset@maths.ox.ac.uk, $^2$cquigley@physics.utoronto.ca, $^3$eirik.svanes@kcl.ac.uk}, Callum Quigley$^{b\,2}$, Eirik Eik Svanes$^{cdef\,3}$}\non\\

\vskip 0.2cm

{\it\small $^{a}$Mathematical Institute, University of Oxford,
Andrew Wiles Building, Oxford, OX2 6GG, UK} \non\\ \vskip 0.2cm

{\it\small $^{b}$Department of Physics, University of Toronto,
Toronto ON, M5S 1A7, Canada} \non\\ \vskip 0.2cm

{\it\small $^{c}$Department of Physics, King's College London, London, WC2R 2LS, UK}
\non\\ \vskip 0.2cm

{\it\small $^{d}$The Abdus Salam International Centre for Theoretical Physics, 34151 Trieste, Italy}
\non\\ \vskip 0.2cm

{\it\small $^{e}$Sorbonne Universit\'es, CNRS, LPTHE, UPMC Paris 06, UMR 7589,
75005 Paris, France}  \non\\ \vskip 0.2cm

{\it\small $^{f}$Institut Lagrange de Paris, 75014 Paris, France}

\end{center}

\begin{abstract}
Recently, the infinitesimal moduli space of heterotic $G_2$ compactifications was described in supergravity and related to the cohomology of a target space differential. In this paper we identify the marginal deformations of the corresponding heterotic nonlinear sigma model with cohomology classes of a worldsheet BRST operator. This BRST operator is nilpotent if and only if the target space geometry satisfies the heterotic supersymmetry conditions. We relate this to the supergravity approach by showing that the corresponding cohomologies are indeed isomorphic. We work at tree-level in $\alpha'$ perturbation theory and study general geometries, in particular with non-vanishing torsion.
\end{abstract}

\end{titlepage}

\newpage

\tableofcontents

\section{Introduction}

Supersymmetric compactifications of the heterotic string have a rich history, starting with internal Calabi-Yau manifolds and the standard embedding of the spin connection in the gauge connection \cite{Candelas1985a}. In recent years, substantial efforts have gone into improving the understanding of heterotic vacua beyond this restricted case: dimensions other than six being studied, and non-trivial torsion being allowed. Less is known about such compactifications due to the appearance of torsion \cite{Strominger1986, Hull1986SS}, and more exotic holonomy groups such a $G_2$ and $Spin(7)$ in dimensions other than six. In particular, it is not known whether higher order $\alpha'$ corrections can spoil spacetime supersymmetry and worldsheet conformal invariance, in contrast to what is known about $(2,2)$ and $(1,1)$ models \cite{nemeschansky1986conformal, witten1986new, Becker:2014rea}. We do not believe this to be the case, since it is expected that the recent recasting of heterotic geometry in terms of holomorphic structures, extension sequences and Courant algebroids \cite{Anderson:2014xha, delaOssa:2014cia, Garcia-Fernandez:2015hja, Clarke:2016qtg, delaOssa:2016ivz, delaOssa:2017pqy, delaOssa:2017gjq} will survive to higher orders, potentially modulo certain field redefinitions.

In this work, we consider a seven-dimensional internal space,
\begin{equation} \label{eq:3+7}
{\cal M}_{9+1}={\cal M}_{2+1}\times Y,
\end{equation}
where $\mathcal{N}=1$ supersymmetry in the external $(2+1)$-dimensional spacetime requires the existence of a $G_2$-structure on $Y$. Though this situation might be of less interest for particle physics phenomenology, it has applications in mathematics. Indeed heterotic geometries of this type are equipped with an instanton bundle, and recently the study of such bundles over manifolds of exceptional structure has become a hot topic. See e.g. \cite{donaldson1998gauge, lewis1999spin, brendle2003construction, donaldson2009gauge, sa2009instantons, Harland:2009yu, Gemmer:2011cp, walpuski2013g2, walpuski2017spin, walpuski2015ex, Haupt:2015wdq, sa2015g2, sa2015g22, menet2015construction, Joyce:2016fij, Haydys2015, Munoz:2016rgc} for a representative but far from exhaustive list of works. Due to the exotic nature of the structures involved, one has much less control than in the corresponding Calabi-Yau setting. New approaches are therefore welcome. In particular, perhaps a worldsheet approach can shed some light on open problems in this setting.

Meanwhile, a lesson revealed by the $SU(3)$ holonomy case is the distinguished role played by infinitesimal deformations of the compactification. Moduli indeed occur as massless fields in the spacetime effective theory and are relevant to topological string theory and mirror symmetry on Calabi-Yau manifolds. Lately, focus has shifted to the study of moduli spaces for more general compactifications, including torsion and in diverse dimensions, mostly from a supergravity point of view \cite{Anderson:2014xha, delaOssa:2014cia, Garcia-Fernandez:2015hja, Clarke:2016qtg, delaOssa:2016ivz, delaOssa:2017pqy, Becker:2006xp, Anderson:2010mh, Anderson:2011cza, Cicoli:2013rwa}, but also in terms of nonlinear sigma models on the worldsheet \cite{Melnikov2011}. In this note, we initiate a study of infinitesimal moduli of general heterotic $G_2$ compactifications from the worldsheet point of view. Using a $(0,1)$ nonlinear sigma model (NLSM) description, we obtain constraints satisfied by the moduli which we then relate to the cohomology of a target space differential $\check{\mathcal{D}}$. We do not rely on the standard embedding nor on any artificial assumptions on the flux. We work at tree-level in the $\alpha'$ expansion and, in this limit, our results agree with those recently obtained in \cite{Clarke:2016qtg, delaOssa:2016ivz, delaOssa:2017pqy} using the supergravity perspective.

Although our leading order result is already known from these works, our method can likely be extended to make exact worldsheet statements via topological twisting, in analogy the results of \cite{Adams:2005tc,Donagi:2011uz,Donagi:2011va} in the $(0,2)$ setting.
Our arguments borrow from the analysis of the superconformal algebra \cite{Shatashvili1995} associated to type II compactifications on manifolds of $G_2$ holonomy. In particular we relate our differential $\check{\mathcal{D}}$ to a certain BRST operator conjectured alongside a topological twist for these theories \cite{Shatashvili1995,DeBoer2005}. Consistency of our results with the supergravity approach tends to support the idea that the superconformal algebra not only holds, but may perhaps also be twisted and used to probe $G_2$-structures deep in the stringy regime. We hope to report on these fascinating (quasi) topological models in the near future.

In the rest of the introduction, we make more precise the target space geometry assumed in our calculation. We then introduce in section~\ref{sec:G2CFT} the necessary ideas from conformal field theory (CFT) on $G_2$ targets. Our calculation is presented in section~\ref{sec:moduli} and in section~\ref{sec:Differential} we connect with target space cohomology. We comment on future prospects in section~\ref{sec:Conclusion}.

\subsection{Geometry of heterotic $G_2$ systems} \label{sec:G2Geom}

We begin by recalling that a $G_2$-structure manifold can be defined as a doublet $(Y,\varphi)$, where $Y$ is seven-dimensional and $\varphi$ is a three-form,
\begin{equation}
\varphi\in\Omega^3(Y),
\end{equation}
satisfying a \textit{positivity} condition. Such a three-form gives rise to a metric, and positivity implies that this metric is positive-definite. A canonical four-form can be constructed as $\psi=*\varphi$ using this metric to define the Hodge star. For more details, see e.g. \cite{Joyce2000,Grigorian2009a}.

Preservation of spacetime $\mathcal{N}=1$ supersymmetry for the ansatz \eqref{eq:3+7} implies the existence of a globally defined real spinor $\eta$ on $Y$ . From this spinor, a $G_2$-structure is defined by $\varphi_{ijk}=-i\eta^T\gamma_{ijk}\eta$, where $\gamma_i$ are the seven-dimensional Dirac gamma matrices. In addition to this topological requirement, the geometry should also satisfy some differential conditions. In particular, the gravitino equation requires\footnote{In this paper, we use the opposite sign convention for $\nabla^{\pm}$ as compared to \cite{delaOssa:2017pqy}, as it is more conventional in the worldsheet literature. In particular, the connection we call $\nabla^+$ is denoted $\d_\theta$ in \cite{delaOssa:2017pqy}.}
\begin{equation}
\label{eq:Spinor}
\nabla^-_i\eta=0,
\end{equation}
or equivalently
\begin{equation}
\label{eq:Parallell}
\nabla^-_i\varphi_{jk\ell}=0.
\end{equation}
The connection symbols of $\nabla^-$ are defined by
\begin{equation} \label{eqn:torsionconnections}
{\Gamma^{\mp}_{ij}}^k={\Gamma_{ij}}^k\mp\tfrac12{H_{ij}}^k={\Gamma^{\pm}_{ji}}^k,
\end{equation}
where $\Gamma$ denote the symbols of the Levi-Civita connection and $H=\d B$ is the flux. 

It is worth making a couple of remarks concerning such heterotic $G_2$-structures. The first remark is that \eqref{eq:Parallell} {\it uniquely} determines the flux $H$ in terms of the intrinsic torsion classes of the $G_2$-structure. The \emph{torsion classes} of a $G_2$-structure are differential forms $\tau_p$ of degree $p=0,1,2,3$ arising in the decomposition of $\d\varphi$ and $\d\psi$ into irreducible representations of $G_2$. Using \eqref{eq:Parallell}, the following identification can be made:
\begin{equation}
H=-\frac{1}{6}\tau_0\varphi+\tau_1\lrcorner\psi+\tau_3.
\end{equation}
A $G_2$-structure satisfying \eqref{eq:Parallell} is moreover of the {\it integrable} type which means that the torsion class $\tau_2$ vanishes. This implies that the geometry has certain interesting features which we will return to in section~\ref{sec:Differential}. Since $\tau_2=0$, we see that the case $H=0$ is equivalent to trivial torsion, which is in turn equivalent to the Levi-Civita connection having holonomy contained in $G_2$. For more details on $G_2$-structures and intrinsic torsion, see e.g. \cite{Friedrich2003, Gauntlett:2003cy, Gran2005, Lukas2011, Gray2012} and references therein.

Next, we comment on the instanton conditions of vector bundles over $Y$. For a gauge bundle $V\rightarrow Y$, the $G_2$ instanton condition
\begin{equation}
\label{eq:InstantonF}
F\wedge\psi=0
\end{equation}
is a consequence of supersymmetry. Here $F$ is the curvature of the gauge field $A$. We then remark that, at tree-level in $\alpha'$ the connection $\Gamma^+$ on $TY$ defined by \eqref{eqn:torsionconnections} satisfies the same instanton equation \cite{Hull1986SS, Hull1986, Bergshoeff1989}:
\begin{equation}
\label{eq:InstantonR}
R^+\wedge\psi=0.
\end{equation}
This can be deduce from \eqref{eq:Parallell} and by noting that
\begin{equation}
\label{eq:RpRm}
R^+_{ijpq} - R^-_{pqij} = (\d H)_{ijpq}\:,
\end{equation}
where $\d H= O(\a')$ can be set to zero in our analysis. With this assumption, the following statements are equivalent:
\begin{itemize}
  \item The connection $\nabla^-$ has $G_2$ holonomy;
  \item The last two indices of $R^-$ transform in the adjoint (${\bf 14}$) of $G_2$;
  \item The first two indices $R^+$ transform in the adjoint (${\bf 14}$) of $G_2$;
  \item The curvature $R^+$ satisfies an instanton condition analogous to \eqref{eq:InstantonR}.
\end{itemize}
Note also that the connection $\Gamma^+$ is the one appearing naturally in the heterotic Bianchi identity
\begin{equation}
\d H=\tfrac{\alpha'}{4}\left(\tr\:R^+\wedge R^+-\tr\:F\wedge F\right)+O(\alpha'^2)\:,
\end{equation}
when the leading $\alpha'$ corrections are introduced. It should be stressed that $R^+$ fails to satisfy the instanton condition when such $\alpha'$ corrections are included. This observation is however crucial for understanding the heterotic moduli problem at $O(\alpha')$ \cite{delaOssa:2017pqy}.

\section{$G_2$ CFTs} \label{sec:G2CFT}
We now come to studying CFTs where the target space has a $G_2$ structure. Minimally supersymmetric $(0,1)$ NLSMs may be formulated for any target manifold. As we presently discuss, restricting to targets with a $G_2$-structure enhances the worldsheet superconformal symmetry to what we will call the \textit{$G_2$ algebra}. This algebra was first written down in the context of type II compactifications in \cite{Shatashvili1995}, although it appeared previously in \cite{Blumenhagen1992} (see also \cite{Figueroa-OFarrill1997, DeBoer2005, Howe1993}).  However, it received significantly less attention in the chiral $(0,1)$ setting of the heterotic string. It was shown recently in \cite{Melnikov2017a} that this algebra is required by preservation of minimal ${\cal N}=1$ supersymmetry in heterotic compactifications with a Minkowski spacetime ${\cal M}_{2+1}$.

We already mentioned that a $G_2$-structure on $Y$ can be characterized as the existence of a positive three-form $\varphi$. The positivity condition ensures the existence of local frames $\{e^i\}$ for the cotangent bundle of $Y$ such that the three-form can be written as
\begin{equation}
\varphi=e^{125} + e^{136} + e^{147} - e^{237} + e^{246} - e^{345} + e^{567},
\end{equation}
where we denote $e^{ijk}=e^i\wedge e^j\wedge e^k$ \cite{Joyce2000}. With $e^i$ substituted by $\d x^i$ in this expression, where $x^i$ are coordinates on $\mathbb{R}^7$, this defines a $G_2$-structure $\varphi_0$ on $\mathbb{R}^7$, regarded as the imaginary octonions (see e.g. \cite{Grigorian2009a} for further details). Thus, simply stated, the positivity condition locally identifies $(Y,\varphi)$ with $(\mathbb{R}^7,\varphi_0)$.

This property heuristically suggests that an extended superconformal algebra derived using free fields, corresponding to flat target space directions, will survive even for non-flat targets with a general $G_2$-structure. This idea was exploited successfully in \cite{Shatashvili1995}, under the assumption that the $G_2$-structure is torsion-free. We emphasize that this restriction, often made in the literature, is not necessary in the argument and indeed our moduli calculation will hold even for nontrivial torsion.

Correlated with the positive three-form, the $G_2$ algebra contains a chiral weight-$3/2$ Virasoro primary operator $\bar{\Phi}$ which acts as a current for the additional symmetry. The other important operators are the $(0,1)$ generators $\bar{T}$ and $\bar{G}$, a weight-$2$ field denoted $\bar{X}$, and the superpartners of $\bar{\Phi}$ and $\bar{X}$. We refer to \cite{Shatashvili1995,Figueroa-OFarrill1997} for further details and the explicit operator product expansions (OPE).

Defining $\bar{T}_I=-\bar{X}/5$ and $\bar{G}_I=i\bar{\Phi}/\sqrt{15}$, the relevant OPE relations rescale into those of ${\cal N}=1$ Virasoro, with central charge $7/10$. This subsector is isomorphic to the tri-critical Ising model and plays a role analogous to the $U(1)_{\text{R}}$ current algebra of $(0,2)$ CFTs~\cite{Shatashvili1995,DeBoer2005}. Although it is not orthogonal to the original superconformal symmetry, it commutes with the ``remaining symmetry'': defining $\bar{T}_r=\bar{T}-\bar{T}_I$, one finds the OPE
\begin{equation}
\bar{T}_I\bar{T}_r=\text{regular}.
\end{equation}
This relation ensures that the weights $(\bar{h}_I,\bar{h}_r)$ with respect to $\bar{T}_I$ and $\bar{T}_r$ can be used to label the spectrum of the $G_2$ algebra.

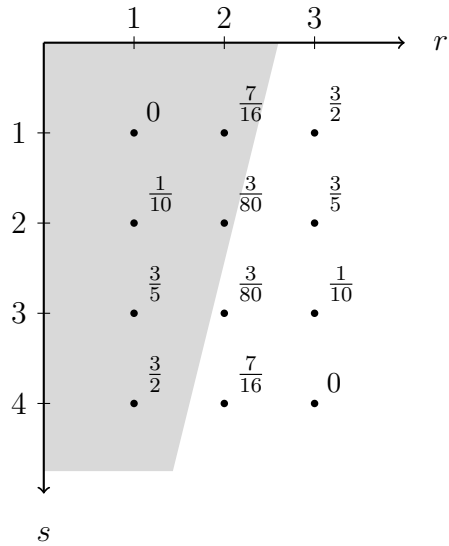
\begin{figure}[htbp]
\begin{center}
\begin{tikzpicture}[scale=1.2,label distance=-1pt]
\fill[fill=gray!30] (-2,0) -- (0.6,0) -- (-0.57,-4.75) -- (-2,-4.75);
\draw[line width=0.8pt,->] (-2,0) -- (2,0);
\draw[line width=0.8pt,->] (-2,0) -- (-2,-5);
\foreach \x in {1,2,3}
\draw (\x - 2, 2pt) -- (\x -2 , -2pt) node[above=4pt] {$\x$};
\foreach \y in {1,2,3,4}
\draw (-2cm + 2pt, -\y ) -- (-2cm - 2pt,-\y) node[anchor=east] {$\y$};
\node at (2.4,0) {$r$};
\node at (-2,-5.45) {$s$};

\weight $0$ at (-1,-1)
\weight $\tfrac{7}{16}$ at (0,-1)
\weight $\tfrac{3}{2}$ at (1,-1)
\weight $\tfrac{1}{10}$ at (-1,-2)
\weight $\tfrac{3}{80}$ at (0,-2)
\weight $\tfrac{3}{5}$ at (1,-2)
\weight $\tfrac{3}{5}$ at (-1,-3)
\weight $\tfrac{3}{80}$ at (0,-3)
\weight $\tfrac{1}{10}$ at (1,-3)
\weight $\tfrac{3}{2}$ at (-1,-4)
\weight $\tfrac{7}{16}$ at (0,-4)
\weight $0$ at (1,-4)

\end{tikzpicture}
\end{center}
\vspace{-0.5cm}\caption{Kac table of the tri-critical Ising minimal model. 
}\label{fig:kac}
\end{figure}

Several properties of the $G_2$ algebra follow from the tri-critical Ising subsector, so we briefly recall some of the prominent features of this model. It belongs to the series of unitary minimal models ($m=4$ in the conventions of \cite{DiFrancesco} used below). As for any such CFTs, it has finitely many conformal families. They are labelled by a pair of integers $\{r,s\}$, with $1\leq r\leq m-1$, $1\leq s\leq m$, and subject to the identifications $\{r,s\}\sim \{m-r , m+1-s\}$. They may be conveniently organised into a lattice in the $r,s$-plane, the \textit{Kac table} (figure~\ref{fig:kac}). Consistently with the identifications, two copies of each conformal family appear in the Kac table. On each site of this lattice, we record the weight of the primary for the corresponding conformal family.

Conformal families of minimal models obey specific fusion rules. The most relevant for us will be
\begin{equation}
\{1,2\}\times\{r,s\}=\{r,s-1\}\oplus\{r,s+1\},
\end{equation}
which states that the OPE between any field in $\{1,2\}$ and any field in $\{r,s\}$ only involves fields in the families immediately above and below $\{r,s\}$ in the Kac table.\footnote{In the case where $\{r,s\}$ is on the boundary, then only the existing neighbour contributes.} Any field in $\{1,2\}$ thus induces two maps obtained by restricting the OPE to either of the neighbouring conformal families. Accordingly, $\{1,2\}$ admits a \textit{conformal blocks} decomposition
\begin{equation}
\{1,2\}=\{1,2\}^{\uparrow}\oplus\{1,2\}^{\downarrow},
\end{equation}
with the following actions:
\begin{align}
\{1,2\}^{\uparrow}&\times\{r,s\}=\{r,s-1\}\\
\{1,2\}^{\downarrow}&\times\{r,s\}=\{r,s+1\}.
\end{align}
The arrows convey the idea of moving up or down in the Kac table. For consistency of these notations, we must restrict $\{r,s\}$ to the shaded half of figure~\ref{fig:kac}.

\subsection{BRST operator} \label{sec:forms}

Returning to the full $G_2$ algebra, the conformal block decomposition will apply to any field whose tri-critical Ising piece lies fully within the family $\{1,2\}$. In particular, this is the case for the $(0,1)$ supersymmetry current \cite{Shatashvili1995,DeBoer2005}:
\ba
\bar G(\bar z) = \bar G^\uparrow(\bar z) + \bar G^\downarrow (\bar z).
\ea
The decomposition also carries over to Laurent modes, and following \cite{DeBoer2005} this allows us to define a BRST operator for the theory as
\ba \label{eq:QBRST}
Q_{\text{BRST}} = \bar Q^\downarrow,
\ea
where $\bar{Q}=\bar{G}_{-1/2}$ is the supersymmetry charge. In \cite{Shatashvili1995,DeBoer2005}, evidence was gathered to support the existence of a topological twist for $G_2$ theories, similar the A and B twists of $(2,2)$ models \cite{Witten1988}. The BRST operator \eqref{eq:QBRST} was proposed in this context, specifically for $(1,1)$ models with $G_2$ holonomy targets. Physical observables of the twisted theory are those that lie in BRST cohomology, including marginal deformations. The BRST operator is key to our calculation in section~\ref{sec:moduli}.

For our purposes, it will be sufficient to act with the BRST operator on superfields of the form
\begin{equation} \label{eq:form}
\Xi^p=\Xi_{i_1i_2\ldots i_p}(\Phi)\bar{D}\Phi^{i_1}\bar{D}\Phi^{i_2}\ldots\bar{D}\Phi^{i_p},
\end{equation}
where our superspace notations will be clarified shortly in section~\ref{sec:moduli}. In the large radius limit, superfields of this form are in one-to-one correspondence with differential $p$-forms on target space. A standard fact about $G_2$-structures is that differential forms decompose into subspaces according to irreducible $G_2$ representations. We may thus similarly decompose superfields $\Xi^p$ using projection operators available in the mathematical literature. For instance, a two-form superfield may be written as
\begin{equation}
\Xi^2=[(\pi_{\bf 7}^2)^{ij}_{mn}+(\pi_{\bf 14}^2)^{ij}_{mn}]\Xi_{ij}\bar{D}\Phi^{m}\bar{D}\Phi^{n},
\end{equation}
where the projectors 
\begin{equation}
(\pi_{\bf 7}^2)^{ij}_{mn}=6\,\varphi^{ij}{}_k\varphi^k{}_{mn},\quad \textrm{and},\quad (\pi_{\bf 14}^2)^{ij}_{mn}=\delta^{i}_{[m}\delta^j_{n]}-(\pi_{\bf 7}^2)^{ij}_{mn}
\end{equation}
are given in terms of the $G_2$-structure three-form $\varphi$.

The BRST action on $\Xi^p$ is determined by its weights. The right-moving weight is easily read off to be $\bar{h}=p/2$, while the tri-critical Ising weight can be obtained from the OPE with $\bar{X}$. As explained in \cite{DeBoer2005} (for the leading component of $\Xi^p$), it is possible to establish in this way a correspondence between irreducible $p$-forms and states $\ket{\bar{h}_I,\bar{h}_r=\bar{h}-\bar{h}_I}$ of the $G_2$ CFT Hilbert space. We state the final result for low degrees in table~\ref{tab:forms-states}.

\begin{table}
\begin{center}
\renewcommand\arraystretch{1}
\begin{tabular}{ c | c | c | c | c }
    Degree & \multicolumn{4}{c}{Irreps} \\
    $p$ & {\bf 1} & {\bf 7} & {\bf 14} & {\bf 27} \\
    \hline
    $0$ & \cellcolor{gray!30} $\ket{0,0}$ & - & - & - \\
    \hline
    $1$ & - & \cellcolor{gray!30} $\ket{\tfrac{1}{10},\tfrac{2}{5}}$ & - & - \\
    \hline
    $2$ & - & \cellcolor{gray!30} $\ket{\tfrac{3}{5},\tfrac{2}{5}}$ & $\ket{0,1}$ & - \\
    \hline
    $3$ & \cellcolor{gray!30} $\ket{\tfrac{3}{2},0}$ & $\ket{\tfrac{11}{10},\tfrac{2}{5}}$ & - & $\ket{\tfrac{1}{10},\tfrac{7}{5}}$
\end{tabular}
\caption{Correspondence between differential forms and $G_2$ CFT states}
\label{tab:forms-states}
\end{center}
\end{table}

The key feature of this table is that the weights of the tri-critical Ising primaries appear in almost all entries. Crucially, the tri-critical weights in the shaded boxes match the first column ($r=1$) of the Kac table in figure~\ref{fig:kac}. Moreover, we see that the form degree $p$ increases in the direction of $\downarrow$ from figure~\ref{fig:kac}. This correspondence between irreducible superfields and differential forms will allow us in section~\ref{sec:moduli} to distinguish the $\downarrow$ conformal block, relevant to the BRST action, from the image of $\bar{Q}^{\uparrow}$ by projecting onto the appropriate $G_2$ irreducible representations.

\section{Moduli of the $G_2$ NLSM} \label{sec:moduli}

In this section, we obtain the infinitesimal moduli space of heterotic $G_2$ systems as the cohomology of the BRST operator $\bar G^\downarrow_{-1/2}$ using a $(0,1)$ NLSM formulation. After setting up our NLSM notations, we proceed with our calculation and verify the nilpotency of $Q_{\text{BRST}}$ on trivial deformations.

\subsection{Review of $(0,1)$ NLSMs}

Working in the conventions of~\cite{Polchinski2}, bosonic and Fermi $(0,1)$ superfields have the following component expansions:
\ba
\F = \f + i\bar\th \j,\qquad \La = \l +\bar\th F.
\ea
We assign weights $(h,\bar h)=(0,0)$ and $(\tfrac12, 0)$ to the $\F$ and $\La$ superfields, respectively. The supercharge and superspace derivative, both of weight $(0,\tfrac12)$, are respectively defined as
\ba
\bar Q = \del_{\bar\th} -\bar\th \bar\del,\qquad \bar D = \del_{\bar\th} +\bar\th \bar\del
\ea
and satisfy $-\bar Q^2 =\bar D^2 = \bar\del$. The general $(0,1)$ NLSM takes the form
\be\label{eqn:nlsm}
S=\frac{1}{4\pi}\int \d^2z \d\bar\th \left(\left[G_{ij}(\F)+B_{ij}(\F)\right]\del\F^i \bar D\F^j  -\La^A \bar D\La^A + i A_i^{AB}(\F) \La^A\bar D\F^i\La^B\right).
\ee
Interpreting the bosonic fields $\phi$ as local coordinates on target space $Y$, the functional couplings $G$, $B$ and $A$ locally specify the metric, $B$-field and gauge field respectively. Up to a possible shift in the gauge field, we can always choose canonical kinetic terms for the Fermi superfields without loss of generality. We may also retain only the antisymmetric part of the gauge field as any symmetric part would drop from~\eqref{eqn:nlsm}.

The equations of motion for the superfields are easily derived:
\ba
&\bar D\del\F^i +\G^{-i}_{jk}\del \F^j \bar D\F^k  +\tfrac{i}{2}  G^{ij}F_{jk}^{AB}\bar D\F^k \La^A\La^B =0, \label{eqn:eomPhi}\\
& \bar D\La^A - i A_i^{AB}(\F) \bar D\F^i\La^B=0. \label{eqn:eomLambda}
\ea
Here, we defined the gauge field strength in the usual way,
\ba
F_{ij}^{AB}=2\left(\partial_{[i} A_{j]}^{AB}-iA_{[i}^{AC}A_{j]}^{CB}\right),
\ea
and the torsionful connections $\G^{\pm}$ as in \eqref{eqn:torsionconnections}.

\subsection{Marginal couplings and BRST cohomology}

We now consider deformations of the action \eqref{eqn:nlsm} of the form
\ba
\dd S = \frac{1}{4\pi} \int \d^2z \d\bar\th\ \O,
\ea
where, given the weights of the various $(0,1)$ superfields and operators, the most general (classically) marginal operator we can deform the NLSM by is
\ba
\O = \dd M_{ij}(\Phi) \del \F^i\bar D\F^j + i\dd A_i^{AB}(\Phi) \La^A \bar D\F^i \La^B. \label{eqn:deformations}
\ea
Here, $\delta M$ and $\delta A$ are general functions of the superfields $\Phi$ allowed on dimensional grounds. The notation is chosen consistently with the NLSM couplings. For instance we can identify the symmetric and anti-symmetric parts of $\dd M$ with deformations of the metric and $B$-field: $\dd M_{ij} = \dd M_{(ij)} +\dd M_{[ij]}  = \dd G_{ij} +\dd B_{ij}$. We do not consider a deformation of the Fermi fields' metric, as this can be reabsorbed into a deformation of the gauge field.

To identify the moduli of the theory, we seek deformations that are BRST closed up to the equations of motion \eqref{eqn:eomPhi}--\eqref{eqn:eomLambda} of the unperturbed NLSM. Working on-shell in this way is not uncommon, as evidenced by similar studies on moduli of $(0,2)$ \cite{Melnikov2011} and $4d$ $\mathcal{N}=1$ \cite{Beasley2006} NLSMs. Proceeding, we notice that $\bar Q =\del_{\bar \th}- \bar\th \bar\del $ only differs from $\bar D$ by a total derivative which drops out of the integrated action. Thus, we consider
\ba \label{eq:DO}
\bar D\O = \dd & M_{ij}\del \F^i \bar\del \F^j + \dd M_{i[j,k]} \bar D\F^k \del \F^i \bar D\F^j + \dd M_{ij} \bar D\del \F^i \bar D\F^j \\
&+ 2i\bar D\La^A \dd A^{AB}_i \bar D \F^i \La^B - i\La^A\dd A^{AB}_{[j,i]}\bar D\F^i \bar D\F^j \La^B - i\La^A \dd A^{AB}_i \bar\del \F^i \La^B. \non
\ea
Following \cite{DeBoer2005}, the first and last terms have $\bar h_I=0$ and can be considered as zero-form superfields in the sense of section~\ref{sec:forms},\footnote{Strictly speaking, they are not exactly of the form \eqref{eq:form}, but left-moving Fermi superfields do not alter right-moving weights and ordinary derivatives of the bosonic superfields may also be ignored \cite{DeBoer2005}.} while the original deformation $\cal O$ corresponds to a one-form superfield. As explained in section~\ref{sec:forms}, we can read off from table~\ref{tab:forms-states} that the first and last terms of \eqref{eq:DO} lie in the image of $\bar G^\uparrow_{-1/2}$, and can thus be dropped. Up to the equations of motion, the remaining terms are then
\ba
\left(\dd M_{i[j,k]} -\G^{-\ell}_{i[k|}\dd M_{\ell|j]}\right)\del\F^i \bar D\F^k \bar D\F^j
 -i\left(D_{[i}\dd A_{j]}^{AB} + \tfrac12 F^{AB}_{k[i}\dd M^k{}_{j]} \right) \bar D\F^i \bar D\F^j \La^A\La^B \non,
\ea
where $D_i$ is the gauge-covariant derivative, acting as
\begin{equation}
D_i(\delta A)_{jAB}=\partial_i(\delta A)_{jAB}-i[A,\delta A]_{AB}.
\end{equation}

To isolate further the image of $Q_{\text{BRST}}$, we must project out the remaining terms with $\bar h_I=0$, as they lie in the ${\bf 14}$ representation of $G_2$. We retain only components in the ${\bf 7}$ representation, which we may select for using the projection operator $(\pi^2_{\bf 7})^{ij}_{mn} = 6\varphi^{ij}{}_k\varphi^k{}_{mn}$. Therefore, the action of the BRST operator on $\O$ is given by:
\ba
\label{eqn:Qclosed}
Q_{\text{BRST}}\O =& \left[\left(\dd M_{k[j,i]} -\G^{-\ell}_{k[i|}\dd M_{\ell|j]}\right)\del\F^i \right.\\
&-i\left(D_{[i}\dd A_{j]}^{AB} + \tfrac12 F^{AB}_{k[i}\dd M^k{}_{j]} \right)\La^A\La^B \Big]\times(\pi^2_{\bf 7})^{ij}_{mn}\bar D\F^m \bar D\F^n\:. \non
\ea
Thus we find that the BRST closed deformations of the action must satisfy the constraints
\ba
\label{eq:MClosure}
&\varphi_m{}^{ij}\left(\dd M_{k[j,i]} -\G^{+\ell}_{[i|k}\dd M_{\ell|j]}\right) = 0 , \\
\label{eq:G2Atiyah}
& \varphi_m{}^{ij}\left(D_{[i}\dd A_{j]}^{AB} + \tfrac12 F^{AB}_{k[i}\dd M^k{}_{j]} \right) =0.
\ea
Note that we have used the index swapping relation~\C{eqn:torsionconnections}\ to convert the $\G^-$ connection appearing in~\C{eqn:Qclosed}\ to $\G^+$ in~\C{eq:MClosure}. These are precisely the (tree-level) relations recently derived from spacetime considerations in~\cite{delaOssa:2016ivz,delaOssa:2017pqy}. The first condition was derived in \cite{DeBoer2005} for type II models in the special case $H=0$. The second condition is of course unique to the heterotic string.

Since our interest is in the cohomology of $Q_{\text{BRST}}$, we must also identify
\ba
\O \sim \O + Q_{\text{BRST}}\O',
\ea
where $\O'$ must have weights $(1,0)$. The most general $\O'$ is therefore of the form
\ba
\O' = C_i(\F)\del \F^i +i \J^{AB}(\F)\La^A \La^B.
\ea
Working through the action of $Q_{\text{BRST}}$ as above leads us to the following identifications:
\ba
\dd M_{ij} &\sim \dd M_{ij} + \nabla^+_j C_i \label{eqn:Midentify}\\
\dd A_i^{AB} &\sim \dd A_i^{AB} - D_i\J^{AB} - \tfrac12  F_i{}^j{}^{AB}C_j.\label{eqn:Aidentify}
\ea
It is somewhat illuminating to unpack the above relations for $\dd G$ and $\dd B$ separately:
\ba
\dd G_{ij} &\sim \dd G_{ij} + \nabla_{(i}C_{j)} \\
\dd B_{ij} &\sim \dd B_{ij} - \partial_{[i}C_{j]} + \tfrac12 H_{ij}{}^kC_k\\
\dd A_i^{AB} &\sim \dd A_i^{AB} - D_i\J^{AB} - \tfrac12 F_i{}^j{}^{AB}C_j.
\ea
We can clearly identify the redundancies associated with diffeomorphisms, $B$-field transformations, and gauge transformations. The final terms on the second and third lines are associated with the effects of diffeomorphisms on the curvatures $H$ and $F$.

\subsection{$Q_{\rm BRST}^2=0$ and the supersymmetry conditions}
We should also ensure that $Q_{\text{BRST}}$ is actually nilpotent. For our purposes, it will suffice to check this on the operator $\O'$:
\ba
Q_{\text{BRST}}^2 \O' &= Q_{\text{BRST}}\left(\nabla^+_j C_i \del\F^i \bar D\F^j -i \La^A(D_j\J^{AB} -\tfrac12 F^i{}_j{}^{AB}C_i)\bar D \F^j \La^B\right) \\
&= \bar D \F^m \bar D \F^n(\pi^2_{\bf 7})^{ij}_{mn}\left[\left(\del_{[i}\nabla^+_{j]}C_k -\G^{+\ell}_{[i|k}\nabla^+_{|j]}C_\ell\right)\, \del \F^k \right.\non\\
&~~ \left. +i\left(D_{[i}D_{j]}\J^{AB} -\tfrac12 (D_{[i|}(F^k{}_{|j]}{}^{AB}C_k) + F^k{}_{[i}{}^{AB}\nabla^+_{j]} C_k)\right)\La^A\La^B \right] .\non
\ea
The terms on the second line can be written in terms of the curvature, since
\ba \label{eq:CurvatureDef}
\del_{[i}\nabla^+_{j]}C_k -\G^{+m}_{[i|k}\nabla^+_{|j]}C_m
=-\left(\del_{[i}\G^{+\ell}_{j]k} -\G^{+m}_{[i|k}\G^{+\ell}_{|j]m}\right)C_\ell = \tfrac12 R^{+}_{ijk}{}^\ell C_\ell,
\ea
while the last set of terms on the final line vanish, as originally shown in~\cite{delaOssa:2017pqy}. To see the latter statement, we will first simplify those terms, suppressing the gauge indices, as follows:
\ba
D_{[i|}(F_{k|j]}C^k) + F_{k[i}\nabla^+_{j]} C^k &= \left(D_{[i|}F_{k|j]} + F_{\ell[i}\G^{+\ell}_{j]k}\right)C^k \\
&= \tfrac12\left(D_k F_{ij} - \G^{+\ell}_{ik}F_{\ell j} -\G^{+\ell}_{jk}F_{i \ell}\right)C^k \non = \tfrac12(D^-_k F_{ij})C^k,
\ea
where we have imposed the Bianchi identity, $D_{[i}F_{jk]}=0$, used~\C{eqn:torsionconnections}\ to convert $\G^+$ to $\G^-$, and used that $\nabla^+$ is metric-compatible. Also we have introduced $D^- \sim D-\G^-$ for the gauge and spacetime covariant derivative with torsion. Thus we have
\ba
Q_{\text{BRST}}^2\O' = & \left[\tfrac12 R^+_{ijk}{}^\ell C_\ell\, \del \F^k  +\left(2F_{ij}^{AC}\J^{CB} -\tfrac{i}{4} D^-_kF_{ij}^{AB} C^k\right)\La^A\La^B \right] (\pi^2_{\bf 7})^{ij}_{mn}\bar D \F^m \bar D \F^n.
\label{eq:BRSTNil}
\ea
Using \C{eq:Parallell}, and that $R^+$ is an instanton at tree-level in $\alpha'$ as explained in section \ref{sec:G2Geom}, the righthand side vanishes provided the curvature satisfies
\ba\label{eqn:instantons}
\varphi^{ij}{}_k F_{ij}^{AB} = 0\:,
\ea
which is equivalent to the instanton condition \eqref{eq:InstantonF} by Hodge duality. This is consistent with the spacetime description, as discussed in section~\ref{sec:Differential}\ and~\cite{delaOssa:2016ivz}. In particular, note that the instanton condition on $F$, together with~\C{eq:Parallell} implies that the last term of \eqref{eq:BRSTNil} also vanishes:
\ba
({D^-_k}F_{ij})\varphi^{ij}{}_\ell = D^-_{k}(F_{ij}\varphi^{ij}{}_\ell) =0\:.
\ea
We conclude that $Q_{\text{BRST}}^2=0$ provided the supersymmetry equations \eqref{eq:Parallell} and \eqref{eq:InstantonF} are satisfied.

It should be noted that the supersymmetry conditions may also be deduced from the requirement that $Q_{\text{BRST}}^2=0$. Indeed, setting $\J^{CB}=0$ in \eqref{eq:BRSTNil} gives the condition
\begin{equation}
\varphi^{mij}\left(\tfrac12 R^+_{ijk}{}^\ell C_\ell\, \del \F^k  -\tfrac{i}{4} D^-_{k}F_{ij}^{AB} C^k\La^A\La^B\right) = 0\:.
\end{equation}
If this is to be true for generic field values of $\{\Phi^k,\La^A\}$, the two terms must vanish separately. In particular, we must have
\begin{equation}
\varphi^{mij}R^+_{ijk}{}^\ell=0\:,
\end{equation}
which by \eqref{eq:RpRm} implies that the $\nabla^-$ connection has $G_2$ holonomy. It is a general fact that, for a fixed connection $\nabla$ on the tangent bundle, there is a one-to-one correspondence between parallel tensors and tensors fixed by the fiberwise action of the holonomy group ${\rm Hol}(\nabla)$. In the case of $\nabla^-$, $G_2$ holonomy guarantees the existence of a positive three-form invariant under parallel transport, and it is unique up to scale. The only other parallel tensors are the associated metric and four-form, again up to scale. In order to show that $\varphi$ is part of the one-parameter family of parallel three-forms, we note that $\nabla^-$ has totally antisymmetric torsion by construction, and therefore automatically preserves the metric corresponding to $\varphi$. By uniqueness, $\varphi$ is also preserved, yielding \eqref{eq:Parallell}.

Finally, we set $C^k=0$ in \eqref{eq:BRSTNil} which implies the instanton condition for the gauge connection, giving us the required target space supersymmetry conditions.

\subsection{Closure and classical symmetries}
\label{sec:sym}
Here we wish to briefly outline an alternative derivation of \eqref{eq:MClosure} which offers a different interpretation of BRST-closed deformations. In  \cite{howe1993holonomy}, it was shown that the existence of parallel forms on target space imply additional symmetries of the NLSM action. These symmetries are classical precursors to the extended G2 algebra of the exact CFT, discussed in section \ref{sec:G2CFT}.\footnote{The original statement applies to the cases of $(1,1)$ and $(0,1)$ supersymmetry without Fermi superfields, but we expect the generalisation to be straightforward.} For the case at hand, the BPS condition \eqref{eq:Parallell},
\begin{equation}
\label{eq:ParallellHP}
\nabla^-\varphi=0,
\end{equation}
implies the nonlinear symmetry
\begin{equation}
\delta^{(\varphi)}\Phi^i=\epsilon\varphi^i{ }_{jk}(\Phi)\bar{D}\Phi^j\bar{D}\Phi^k.
\end{equation}

Above we considered the marginal deformation $\delta S = \int \mathcal{O}$, realising the substitution $G\rightarrow G+\delta G$ and $B\rightarrow B+\delta B$. Demanding BRST-closure amounts, at the classical level, to preservation of the symmetry
\begin{equation}
\delta^{(\varphi)}\Phi^i=\epsilon(\varphi^i{ }_{jk}+\delta\varphi^i{ }_{jk})(\Phi)\bar{D}\Phi^j\bar{D}\Phi^k,
\end{equation}
where $\delta\varphi$ is a variation of the $3$-form compatible with the variation $\delta G$ of the induced $G_2$ metric. The condition for this transformation to leave $S+\delta S$ invariant is precisely \eqref{eq:ParallellHP} with $\varphi$, $G$ and $B$ substituted by their deformed counterparts. The moduli constraint thus obtained is then
\begin{equation}
0 = (\nabla^-+\delta\nabla^-)(\varphi+\delta\varphi) = \delta (\nabla^-\varphi),
\end{equation}
where we neglected the quadratic term. This is simply the variation of the BPS constraint, which, as explained in \cite{delaOssa:2017pqy}, leads to the constraint in the form \eqref{eq:MClosure} derived above.

\section{The BRST cohomology and a target space differential}
\label{sec:Differential}

We can rephrase the equations \eqref{eq:MClosure}--\eqref{eq:G2Atiyah} in terms of a differential on the target space, which then computes the BRST cohomology. We begin by recalling the canonical $G_2$ cohomology \cite{carrion1998generalization, Fernandez1998}
\begin{equation}
\label{eq:CanonicalComplex}
0\rightarrow\Omega^0(Y)\xrightarrow{\d}\Omega^1(Y)\xrightarrow{\check\d}\Omega^2_{\bf 7}(Y)\xrightarrow{\check\d}\Omega^3_{\bf 1}(Y)\rightarrow0,
\end{equation}
where $\check\d=\pi\circ\d$ and $\pi$ is the projection onto the appropriate $G_2$ representation. This is a differential complex, i.e. $\check\d^2=0$, whenever the $G_2$-structure is integrable, i.e. whenever the torsion class $\tau_2$ vanishes.

Given a general bundle $E$ over $Y$ with a covariant derivative $\d_E$, the complex \eqref{eq:CanonicalComplex} can be extended to a differential complex for bundle-valued forms
\begin{equation}
\label{eq:CanonicalComplexBundle}
0\rightarrow\Omega^0(Y,E)\xrightarrow{\d_E}\Omega^1(Y,E)\xrightarrow{\check\d_E}\Omega^2_{\bf 7}(Y,E)\xrightarrow{\check\d_E}\Omega^3_{\bf 1}(Y,E)\rightarrow0,
\end{equation}
where again $\check\d_E=\pi\circ\d_E$. This is a differential complex if and only if the curvature $F_E$ of the connection $\d_E$ satisfies the instanton condition
\begin{equation}
\label{eq:instgauge}
F_E\wedge \psi=0.
\end{equation}

In particular, we get a differential complex $\check\Omega^*_{\check \d_{A}}(Y,{\rm End}(V))$ on the gauge bundle. Since the connection $\nabla^+$ on the tangent bundle also satisfies the instanton condition \eqref{eq:InstantonR}, we get the differential complex $\check\Omega^*_{\check\d_\theta}(Y,TY)$. Here we have defined index-free versions of our earlier covariant derivatives:
\begin{equation}
\d_A=\d\phi^iD_i\::\quad\Omega^p({\rm End}(V))\rightarrow\Omega^{p+1}({\rm End}(V)),
\end{equation}
while the derivative on the tangent bundle
\begin{equation}
\d_\theta=\d \phi^i\nabla^+_i\::\;\;\;\Omega^p(TY)\rightarrow\Omega^{p+1}(TY),
\end{equation}
is defined such that  $\Gamma^+$ only acts on tangent bundle indices. This is consistent with treating $V$ and $TY$ on the same footing.

In addition, the curvature $F$ of the gauge connection can be used to define an extension connection on the topological sum of bundles
\begin{equation}
{\cal Q}={\rm End}(V)\oplus TY.
\end{equation}
The connection, denoted $\cal D$, is defined as
 \begin{equation}
  {\cal D} =  \left(\,
 \begin{matrix}
  \d_A & {\cal F}
 \\  0 & \d_\theta
 \end{matrix}
 \,\right),\label{eq:calD}
 \end{equation}
where the map
\begin{equation}
{\cal F}\::\quad\Omega^{p}(Y,TY)\rightarrow\Omega^{p+1}(Y,{\rm End}(V))
\end{equation}
is given by
\begin{equation}
\label{eq:mapF}
  {\cal F}(\alpha) = \frac{1}{2}(-1)^{p}\, \alpha^i\wedge F_{ij}\, \d \phi^j,
\end{equation}
for some vector-valued form $\alpha\in \Omega^p(Y, TY)$. It can further be checked that the connection $\cal D$ itself is an instanton connection \cite{delaOssa:2016ivz,delaOssa:2017pqy}. We thus get a short exact sequence of complexes
\begin{equation}
0\rightarrow\check\Omega^*_{\check \d_A}(Y,{\rm End}(V))\xrightarrow{i}\check\Omega^*_{\check{\cal D}}(Y,{\cal Q})\xrightarrow{p}\check\Omega^*_{\check\d_\theta}(Y,TY)\rightarrow0,
\end{equation}
where $i$ and $p$ denote injection and projection respectively.

The marginal deformations can be collected in a doublet $y=(\delta A,\delta M)\in\Omega^1({\cal Q})$, where we have used the metric to raise the first index of $\delta M$ to a tangent bundle index.
We see that the marginality conditions \eqref{eq:MClosure}--\eqref{eq:G2Atiyah} are equivalent to
\begin{equation}
\check{\cal D}y=0.
\end{equation}

Moreover, we see from \eqref{eqn:Midentify}--\eqref{eqn:Aidentify} that BRST exact deformations correspond to $\cal D$-exact one-forms. Hence, modulo symmetries, the infinitesimal marginal deformations coincide with classes
\begin{equation}
\label{eq:Coh}
[y]\in H^1_{\check{\cal D}}(Y,{\cal Q}).
\end{equation}

Using that $\Omega_{\check{\cal D}}(Y,{\cal Q})$ is defined as an extension, we can compute this cohomology using the long exact sequence
\begin{align}
H^0_{\check\d_\theta}(Y,TY)&\xrightarrow{\check{\cal F}} H^1_{\check \d_A}(Y,{\rm End}(V))\xrightarrow{i^*}H^1_{\check{\cal D}}(Y,{\cal Q})\xrightarrow{p^*}H^1_{\check\d_\theta}(Y,TY)\notag\\
&\xrightarrow{\check{\cal F}} H^2_{\check \d_A}(Y,{\rm End}(V))\rightarrow...
\label{eq:les}
\end{align}
where map between cohomologies
\begin{equation}
\check{\cal F}\::\;\;\;H^{p}_{\check\d_\theta}(Y,TY)\rightarrow H^{p+1}_{\check \d_A}(Y,{\rm End}(V))
\end{equation}
is induced from \eqref{eq:mapF} in a very similar fashion to what happens in complex geometry \cite{Atiyah1957}. Indeed, that $\check{\cal F}$ defines a map between cohomologies can be derived from the fact that $\check{\cal D}^2=0$. We can now use the long exact sequence \eqref{eq:les} to compute
\begin{equation}
\label{eq:infModuli}
H^1_{\check{\cal D}}(Y,{\cal Q})\cong\frac{H^1_{\check D}(Y,{\rm End}(V))}{{\rm Im}(\check{\cal F})}\oplus\ker\left(\check{\cal F}\right),
\end{equation}
where $\ker\left(\check{\cal F}\right)\subseteq H^1_{\check\d_\theta}(Y,TY)$. It follows that not all of the marginal deformations $\delta M$ are allowed. Only the ones contained in the kernel of the map $\check{\cal F}$. From a target space perspective these should be understood as deformations of the geometry that preserve the instanton condition on the bundle.

Notice that in \eqref{eq:infModuli} we also mod out the bundle moduli by ${\rm Im}(\check{\cal F})$. This can be understood by considering the symmetry transformations \eqref{eqn:Midentify}--\eqref{eqn:Aidentify}. Recall that diffeomorphisms and gauge transformations of the $B$-field are given in terms of a tangent bundle valued scalar $C^m\in\Omega^0(Y,TY)$. We understand the non-closed ones give rise to the exact forms $\d_\theta C^m$ corresponding to trivial deformations of the metric and $B$-field, and implies that $\delta M$ takes values in the cohomology $H^1_{\check{\d_\theta}}(Y,TY)$. Consider then a closed deformation $\d_\theta C^m$, which does nothing to change $\delta M$ but does change $\delta A$ in a non-trivial way
\begin{equation}
\dd A_i^{AB} \sim \dd A_i^{AB} -\tfrac12  {F_{ij}}^{AB}C^j.
\end{equation}
This precisely corresponds to the modding out by ${\rm Im}(\check{\cal F})$. Of course, for simply connected manifolds of $G_2$ holonomy one has $H^1_{\check{\d_\theta}}(Y,TY)=0$, and so this quotient is trivial. It might however produce a nontrivial effect for more general torsional geometries.

\section{Concluding Remarks} \label{sec:Conclusion}
This paper has been concerned with reproducing from the worldsheet point of view the results of \cite{delaOssa:2016ivz,delaOssa:2017pqy} on the infinitesimal moduli of the heterotic string compactified on geometries with $G_2$-structure. This exercise is part of an ongoing program of understanding the moduli of heterotic geometries compactified on manifolds with exceptional structure. One obvious question to ask next concerns $\alpha'$ corrections. Including such corrections to the $(0,1)$ worldsheet theory tends to be a rather complicated endeavor, partially due to lack of supersymmetry. As a guiding principle, one could use the first order supergravity analysis performed in \cite{delaOssa:2017pqy} to identify the appropriate BRST charge. This is also related to the preservation of symmetries discussion in section \ref{sec:sym}, which on general grounds one also might expect to hold at higher orders. In the supergravity analysis, the infinitesimal moduli were identified as elements of a cohomology similar to \eqref{eq:Coh}, except that the corresponding differential no longer defines ${\rm End}(V)\oplus TY$ as an extension anymore. This fact comes together with the fact that the covarant derivative $\d_\theta$ no longer satisfies the instanton condition due to the non-trivial Bianchi identity at $O(\alpha')$, and implies instead that a covariant derivative of the form
\begin{equation}
 {\cal D} =  \left(\,
  \begin{matrix}
  \d_A &     {\cal F}
 \\  \tilde{\cal F} & \d_\theta
 \end{matrix}
 \, \right)~,
 \end{equation}
 satisfies the instanton condition. The expression for $ \tilde{\cal F}$ is not important for the present discussion, except to say that it is also given in terms of the curvature $F$.\footnote{The expression for $ \tilde{\cal F}$ can be found in \cite{delaOssa:2017pqy}.} This is in agreement with the notion that the heterotic moduli space does not factorize, not even pointwise in this case.

Next, it might be interesting to go beyond infinitesimal deformations and consider the exactly marginal deformations, corresponding to integrable deformations of the geometry from the supergravity point of view. In doing so, we would map out the nature of the deformation algebra of the heterotic $G_2$ system. Is it perhaps a differentially graded Lie algebra, or might there be a more general L-infinity structure behind it? It would be interesting to see what form exactly this structure takes, and how it relates to the nature of the worldsheet CFT.

It would also be interesting to perform a similar calculation in the case of heterotic $Spin(7)$ compatifications to two spacetime dimensions. The algebraic formulation of the SCFT is as well developed as in the $G_2$ case \cite{Shatashvili1995}, and its connection to general heterotic compactifications was also made recently \cite{Melnikov2017}. The Ising model plays a similar role there to the tri-critical Ising subsector for the $G_2$ case. It also participates in a topological twist \cite{Shatashvili1995,DeBoer2005}. It seems to us that the method exhibited in this paper should be applicable to the $Spin(7)$ case with slight modifications. 

Finally, it is hard to avoid speculating on possible quasi-topological sectors for heterotic $G_2$ systems. Similar sectors have been found in the $(0,2)$ setting \cite{Adams:2005tc}, and been used to compute exact results in the worldsheet theory \cite{Donagi:2011uz,Donagi:2011va}. The fact that we reproduce expected results on heterotic moduli gives further credence to the original topological twist and BRST proposals in \cite{Shatashvili1995, DeBoer2005}. It also suggests that $Q_{\text{BRST}}$ continues to apply beyond the realm of $(1,1)$ models and vanishing torsion. Such topological sectors would also be relevant for understanding the nature of any topological theory that might govern the heterotic $G_2$ deformation algebra, and might also help to shed some light on open mathematical problems concerning $G_2$ structure manifolds with instantons bundles \cite{donaldson2009gauge, Joyce:2016fij}.

Topological twists of the $(1,1)$ models are also closely related to the generalisation of mirror symmetry to the $G_2$ setting \cite{Shatashvili1995, Papadopoulos:1995da, Acharya:1996fx, Acharya:1997rh, Braun:2017ryx, Braun:2017uku}. Since there is a notion of heterotic mirror symmetry in terms of quantum sheaf cohomology, see \cite{Melnikov:2012hk, Sharpe:2015vza} with references therein, one might speculate in analogy that a similar generalisation to the $(0,1)$ heterotic $G_2$ setting exists. This all makes the search for (quasi) topological sectors of the heterotic $G_2$ string a very attractive endeavour, and we will return to this topic in future publications.

\section*{Acknowledgements}
We wish to thank Xenia de la Ossa for useful discussions and comments on the draft of this paper. We also thank Dominic Joyce, Spiro Karigiannis and Andreas Braun for helpful comments. EES is grateful for the hospitality of The Abdus Salam International Centre for Theoretical Physics where part of the paper was finalised. MAF is financed by a Reidler scholarship from the Mathematical Institute at the University of Oxford and by a FRQNT doctoral scholarship from the Government of Quebec. CQ was supported by a fellowship from NSERC Canada. The work of EES, partly made within the Labex Ilp (reference Anr-10-Labx-63), was supported by French state funds managed by the Agence nationale de la recherche, as part of the programme Investissements d'avenir under the reference Anr-11-Idex-0004-02. In addition, the work of EES was supported by a grant from the Simons Foundation (\#488569, Bobby Acharya).

\bibliographystyle{amsunsrt-ensp}

\end{document}